# NOTAS METODOLÓGICAS PARA CUBRIR LA ETAPA DE DOCUMENTAR UNA INVESTIGACIÓN

**Jose Daniel Texier R.**

Universidad Nacional Experimental del Táchira (UNET)

Servicio de Difusión de la Creación Intelectual (SeDiCI) de la Universidad Nacional de La Plata

jtexier@unet.edu.ve - dantexier@gmail.com

## 1. Resumen

El proceso de búsqueda de artículos científicos (*papers*) y artículos de revisión (*reviews*) es uno de los pilares del mundo científico, y es realizado por personas que están en la investigación así como también por personas que desean mantenerse actualizados en temas específicos. Scopus (existen otras bases de datos) o Google Scholar son las opciones que se proponen para encontrar artículos, pero se recomienda usar Scopus por su amplia base de datos y su versatilidad en las opciones de búsqueda que ofrece. En el presente trabajo se propone un plan que permite de manera sistemática buscar y mantener los artículos encontrados en forma ordenada, consistente y coherente dentro de un repositorio propio para catalogación y consulta, que servirá para muchas labores como establecer el estado del arte de un tema, formación de personal en un área y/o escribir artículos, entre otros.

Palabras clave: elaboración de artículos, *paper*, *review*, Scopus, repositorios

## 2. Introducción

El seminario doctoral llamado "Publicar y No Morir" permitió entender de una manera más clara el proceso de las publicaciones en el mundo científico, adquiriendo un conjunto de conocimientos para la posible aceptación de artículos en revistas de la corriente principal. Por ello el objetivo principal en este trabajo esta basado en proponer un plan que garantice un artículo de revisión de calidad para su aceptación en revistas del área de estudio del postgrado que se está realizando.

Un artículo científico es un trabajo de investigación (también llamado *paper*) para ser publicado en una revista especializada (apegado a unas normas y arbitrado), en el cual se pretende comunicar un resultado original producto de una investigación y dirigido a terceras personas para transferir el conocimiento generado [1].

En cuanto a un artículo de revisión (*review*), también tiene la modalidad de ser publicado para revistas especializadas (*journals*) y debe ser evaluado (arbitrado). La diferencia con el artículo científico radica que en el *review* se dan a conocer las bases teóricas del tema deseado (el estado del arte) [1]. Algunas revistas exigen en los *reviews* contrastar y añadir opiniones con criterios propios, agregando valor a la simple búsqueda de material de otros autores.

Los *paper*s y los *reviews* contienen estructuras y planteamientos parecidos, el proceso de búsqueda de literatura es similar, así como también la evaluación y el análisis de información; pero sus propósitos son sustancialmente diferentes. El propósito del *review* es dar una visión general de la literatura de un tema y no hay aporte al estado del arte. A diferencia del *paper* que busca un aporte al estado del arte [1].

3. Plan Estratégico

A continuación se describen los pasos sugeridos para construir el estado del arte, es decir, para elaborar un *review*. Medina y otros [2] sugieren una propuesta metodológica para la construcción de un artículo, entonces a partir de este estudio se definen las siguientes fases:

1. Identificación del tema estudio.
2. Selección de de las referencias bibliográficas de acuerdo con el tema de estudio y las fuentes de información.
3. Realización de la búsqueda.
4. Análisis de los resultados.

3.1. Identificación del tema estudio

El tema principal de este trabajo está centrado en la propuesta técnica aprobada por la Facultad de Informática titulada: "Representación de Recursos dentro de una Biblioteca Digital"[3], la cual está relacionada directamente con los repositorios de acceso abierto y las siguientes áreas: representación de recursos, base de datos, preservación digital, metadatos e indexación. El tema de estudio mejorará la calidad del funcionamiento sintáctico y estructural de los repositorios.

3.2. Selección de las referencias bibliográficas de acuerdo con el tema de estudio y las fuentes de información

Después de definir el tema de estudio, se deben seleccionar las fuentes de información para la correcta identificación de las referencias bibliográficas. Las fuentes deben ser bases de datos de documentos científicos, documentos académicos y/o tesis doctorales. Los *proceedings* de eventos y congresos son importantes para el intercambio de ideas y debates de nuevos proyectos, pero el objetivo del trabajo es la elaboración del estado del arte, por ello los *proceedings* no se tomaran en cuenta.

Las bases de datos con las que cuenta la Facultad de Informática, para el 10 de agosto del 2011, y que son referentes al tema de investigación son:
- Scopus (Sciverse / ScienceDirect. Scopus cuenta con 19.003 títulos [4].
- EBSCOhost Source Business Complete tiene 7.962 títulos [4].
- IEEExplore tiene aproximadamente 256 títulos para el 10 de agosto del 2011 [5].
- ISI Web of Knowledge tiene 11.456 títulos [4].

Google Scholar es también una fuente de consulta importante a tomar en cuenta ya que permite dar un panorama general sobre algún tema particular y/o autor de forma gratuita, pero es preferible usar bases de datos de información arbitrada totalmente. Las tesis doctorales brindan una información importante, que suele contener una detallada revisión de la literatura y amplios contenidos sobre un área determinada. Estas tesis y/o documentos académicos se pueden encontrar en los siguientes repositorios:
- Tesis Doctorales en Red: http://www.tdr.cesca.es/.
- Network Digital Libray of Theses and Dissertations (NDLTD).
- Biblioteca Digital de Tesis de Universidades Españolas: http://www.ucm.es/BUCM/tesisdigitales/02.htm.
- Portal de tesis electrónicas: http://www.cybertesis.net/.
- Servicio de Difusión de la Creación Intelectual (SeDiCI): http://sedici.unlp.edu.ar.

Todas las diferentes fuentes nombradas anteriormente permiten acceso a publicación científica, por ello se debe seleccionar una base de datos y un repositorio de documentos académicos que oriente la búsqueda. Nuestra selección se centrará en usar Scopus por sus diversas opciones de búsquedas y por la actualización constante de la base de datos.

Para cada búsqueda filtrada en Scopus que se realice sobre el tema principal y subáreas de estudio, se deben estar analizando los diferentes indicadores bibliométricos que se muestran en los artículos, revistas y/o autores. Por tanto, se seleccionan dos categorías de *journals* que ofrece Scopus, a través del Journal Ranking SCImago:
- Library and Information Sciences que para el 15 de julio tiene 124 *journals*, http://www.scimagojr.com/journalrank.php?category=3309.
- Information Systems que para el 15 de julio tiene 129 *journals*, http://www.scimagojr.com/journalrank.php?category=1710.

Se listan 10 revistas para cada una de las categorías y así poder obtener la corriente principal científica del tema de estudio, resaltando la originalidad, actualidad y calidad de los artículos presentes en ellas. Para obtener las dos listas, se hace una preselección de las revistas (para cada categoría) ya que algunas no tienen afinidad con el tema de estudio, esta preselección la realizan los integrantes del laboratorio de investigación donde me encuentro. El resultado de la preselección se clasifica según su prestigio, que se determina a partir del indicador SJR que ofrece SCImago, para luego seleccionar las 10 mejores:

Categoría *Library and Information Sciences*:
1. Journal of the American Society for Information Science and Technology: Q1, 0.071, 2,68
2. Scientometrics: Q1, 0.066, 1,93
3. Journal of Information Science: Q1    0.055, 1,85
4. Library Hi Tech: Q1, 0.044, 1,08
5. Information Systems Journal: Q1, 0.043, 2,08
6. Information Society: Q1, 0.040, 1,40
7. Reference Librarian: Q2, 0.037, 0,76
8. International Journal of Metadata, Semantics and Ontologies: Q2, 0.037, 1,20
9. Journal of Librarianship and Information Science: Q2, 0.035, 0,89
10. D-Lib Magazine: Q2, 0.035, 0,63

Categoría *Information Systems*:
1. Semantic Web and Information Systems: Q1, 0.082, 30.86
2. Journal of the American Society for Information Science and Technology: Q1, 0.071, 44.82
3. Knowledge and Information Systems: Q1, 0.067, 41.18
4. Information Retrieval: Q1, 0.067, 36.36
5. Journal of the Association of Information Systems: Q1, 0.061, 96.28
6. Journal of Documentation: Q1, 0.056, 50.93
7. Journal of Information Science: Q2, 0.055, 46.00
8. Information and Software Technology: Q2, 0.054, 42.21
9. Journal of Systems and Software: Q2, 0.051, 33.34
10. European Journal of Information Systems: Q2, 0.048, 57.54

Para las dos listas de revistas seleccionadas, es importante destacar que cada una de ellas muestra el cuartil según el SCImago Journal Rank (SJR), luego muestra el valor del indicador SJR y después el promedio de referencias por documento en el 2010.

A continuación se listan las revistas en las que se puede publicar artículos para ser popular, donde popular se entiende por tener artículos que tengan una cantidad de referencias considerables (p.e. por encima del promedio). Las 10 primeras para las dos categorías son:

Categoría *Library and Information Sciences:*
1. Information Systems Journal: Q1, 0.043, 2,08
2. Scientometrics: Q1, 0.066, 1,93
3. Journal of Information Science: Q1, 0.055, 1,85
4. Information Society: Q1, 0.040, 1,40
5. International Journal of Metadata, Semantics and Ontologies: Q2, 0.037, 1,20
6. Library Hi Tech: Q1, 0.044, 1,08
7. Journal of Librarianship and Information Science: Q2, 0.035    0,89
8. Reference Librarian: Q2, 0.037, 0,76
9. Library Trends: Q2, 0.033, 0,71
10. D-Lib Magazine: Q2, 0.035, 0,63

Categoría *Information Systems*:
1. Journal of the Association of Information Systems: Q1, 0.061, 96.28
2. European Journal of Information Systems: Q2, 0.048, 57.54
3. Journal of Documentation: Q1, 0.056, 50.93
4. Journal of Computer Information Systems: Q3, 0.038, 47.34

5. Journal of Information Science: Q2, 0.055, 46.00
6. Journal of the American Society for Information Science and Technology: Q1, 0.071, 44.82
7. Information and Software Technology: Q2, 0.054, 42.21
8. International Journal of Metadata, Semantics and Ontologies: Q3, 0.037, 41.67
9. Knowledge and Information Systems: Q1, 0.067, 41.18
10. Information Retrieval: Q1, 0.067, 36.36

Para las dos listas de revistas seleccionadas, es importante destacar que cada una de ellas muestra el cuartil según el SCImago Journal Rank (SJR), luego muestra el valor del indicador SJR y después el promedio de referencias por documento en el 2010.

Finalmente es importante destacar que se hacen 2 clasificaciones, por prestigio y popularidad, para que el investigador pueda seleccionar donde publicar el artículo; aunque nuestra recomendación es siempre buscar el prestigio del trabajo realizado garantizando la calidad del mismo.

### 3.3. Realización de la búsqueda

La base de datos seleccionada es Scopus, donde la estructura de búsqueda es muy similar a un buscador estándar, con una particularidad que permite ubicar y sugerir fácilmente autores o afiliaciones de los autores existentes a partir del criterio de búsqueda. Una gran ventaja de Scopus es que permite almacenar históricos de las búsquedas y crear diferentes alertas. En la búsqueda realizada, los resultados siempre contendrán falsos positivos (artículos que han sido seleccionados por la búsqueda automática pero que realmente no responden a los objetivos del estudio) y falsos negativos (artículos no detectados por la estrategia de búsqueda establecida pero que son de interés para el estudio)[2], situación que debe tomarse en cuenta para obtener una búsqueda satisfactoria. Esto permite reducir la cantidad de referencias obtenidas, sin perder las que realmente importan.

De igual manera, se debe realizar una comparación entre la búsqueda manual dentro de la revista y la búsqueda automática de Scopus, ya que pueden existir divergencias y siempre se debe garantizar que los resultados son los correctos ya que los buscadores pueden fallar. Luego se deben usar gestores de referencias bibliográficas como RefWorks, Reference Manager, Zotero, End Note, etc., o en su defecto, tener un repositorio propio o gestionar a la antigua usando hojas de calculo. El gestor seleccionado, debe permitir la identificación de los artículos según la evaluación. La clasificaciones de los artículos evaluados serían: seleccionado, falso positivo, falso negativo, dudoso y no seleccionado. Estas clasificaciones permiten tener un control de las lecturas y selecciones realizadas en todo el proceso.

Las búsquedas realizadas y la administración de los artículos evaluados con sus aspectos más importantes serán gestionados a través de un repositorio propio que permita la personalización. Por ello, la herramienta para el repositorio sera DSpace, que permite almacenar el *paper* seleccionado y catalogarlo de acuerdo con los criterios del tema seleccionado. Los aspectos principales a tomar en cuenta para la catalogación del artículo son:
- Título.
- Autor(es).
- Resumen.
- Palabras claves.
- Enunciado del problema de investigación.
- Área y subarea del problema.
- Metodología.
- Resultados.

### 3.4. Análisis de los resultados

A partir de los trabajos seleccionados y catalogados en el repositorio DSpace se comenzará a construir el estado del arte (base conceptual), a través de la evaluación, de acuerdo con las áreas del tema de investigación destacando las siguientes:
- Bibliotecas digitales y repositorios.
- Representación de recursos.
- Preservación digital.
- Indexación.

Siempre es importante encontrar los *seminal papers* de cada área, ya que permite comenzar a trazar toda la ruta de investigación conociendo a los autores más destacados, los laboratorios o centros de investigación referentes y las revistas más significativas, es decir, con mayor prestigio por tener altos valores del SJR.

4. Trabajos a Futuro

La elaboración de esta monografía permitió identificar la importancia de la lectura oportuna de *papers* y *reviews* actualizados, los cuales se pueden ubicar fácilmente a través del servicio ofrecido por Scopus, este proceso estará fuertemente vinculado con el repositorio. Por ello, los trabajos a futuros estarán centrados sobre las personalizaciones realizadas en el repositorio así como también en los diferentes componentes de software desarrollados para fortalecer el método planteado. Los trabajos a futuro son:

- Personalización de los metadatos identificados como los aspectos de los artículos que darán un valor agregado al repositorio y a las búsquedas.
- Desarrollo de software para generación de referencias desde DSpace.
- Ajustar el repositorio de DSpace para no permitir el acceso al público.
- Desarrollo o adaptación de un componente de software que permita la identificación de redundancia de recursos dentro del repositorio.

5. Conclusiones

El desarrollo de este trabajo me permitió establecer un método para diseñar un plan estratégico para la elaboración de *papers* y *reviews*. El método propuesto se centra en primera instancia, en la selección de Scopus para identificar los *papers* y *reviews*, luego se tendrá un repositorio de documentos donde se catalogarán los artículos según los aspectos más importantes. Luego a partir de este repositorio se podrán realizar búsquedas, generar referencias para procesadores de texto como Word, OpenOffice, LibreOffice, Latex, etc., además podrá estar a disposición al público en forma general o parcial.

Este método, se muestra como algo novedoso que servirá como un caso de estudio para las personas que desean tener un almacén de artículos ordenado, consistente y coherente. Por ello, se estará realizando un seguimiento en los avances y pruebas relacionada a este método

6. Referencias Bibliográficas